\documentclass[conference]{IEEEtran}
\IEEEoverridecommandlockouts
\usepackage{cite}
\usepackage{amsmath,amssymb,amsfonts}
\usepackage{graphicx}
\usepackage{textcomp}
\usepackage{xcolor,soul}
\usepackage{cleveref}
\usepackage[nolist,nohyperlinks]{acronym}
\usepackage{verbatim}
\usepackage{algorithm,algpseudocode}
\usepackage{tikz}

\Crefname{figure}{Fig.}{Figs.}

\def\BibTeX{{\rm B\kern-.05em{\sc i\kern-.025em b}\kern-.08em
    T\kern-.1667em\lower.7ex\hbox{E}\kern-.125emX}}
\begin{document}

\title{Dynamic Entanglement Packet Scheduling for Quantum Networks
% \thanks{}
}

\author{
    \IEEEauthorblockN{
    Quang-Phong Tran\IEEEauthorrefmark{1}\IEEEauthorrefmark{2},
    Claudio Cicconetti\IEEEauthorrefmark{2},
    Marco Conti\IEEEauthorrefmark{2},
    Andrea Passarella\IEEEauthorrefmark{2}}
    \IEEEauthorblockA{\IEEEauthorrefmark{1}
    \textit{Department of Information Engineering, University of Pisa}, Pisa, Italy}
    \IEEEauthorblockA{\IEEEauthorrefmark{2}
    \textit{Institute of Informatics and Telematics, National Research Council}, Pisa, Italy\\
    Corresponding author: quangphong.tran@phd.unipi.it}
}

\maketitle

\begin{tikzpicture}[remember picture,overlay]
\node[anchor=south,yshift=10pt] at (current page.south) {\fbox{\parbox{\dimexpr\textwidth-\fboxsep-\fboxrule\relax}{
  \footnotesize{
     \copyright 2026 IEEE.  Personal use of this material is permitted.  Permission from IEEE must be obtained for all other uses, in any current or future media, including reprinting/republishing this material for advertising or promotional purposes, creating new collective works, for resale or redistribution to servers or lists, or reuse of any copyrighted component of this work in other works.
  }
}}};
\end{tikzpicture}%

\begin{abstract}
Sharing entanglement among multiple users remains a central challenge for scalable quantum networks. Recent work proposed an on-demand entanglement packet architecture in which a controller uses a Time Division Multiple Access (TDMA) approach to allocate network resources.
Quantum nodes are assigned a periodic schedule that probabilistically fulfills application requests for end-to-end entanglements. The schedule is recomputed periodically using well-known algorithms, such as \ac{EDF}.
However, a static schedule offers limited flexibility when outcomes are stochastic and arrivals are asynchronous. To overcome this limitation, we propose an online scheduler that dynamically schedules, defers, retries, or drops entanglement distribution reservations.
In our simulations, the dynamic scheduler achieves lower completion time, higher completion ratio, and higher throughput than the static baseline.
Furthermore, when the network is overloaded, the dynamic scheduler continues to construct deadline-feasible schedules and degrades gracefully.
\end{abstract}


\begin{acronym}
    \acro{PGA}{Packet Generation Attempt}
    \acro{EDF}{Earliest Deadline First}
    \acro{E2E}{End-to-End}
    \acro{BSM}{Bell State Measurement}
    \acro{LCM}{Least Common Multiple}
\end{acronym}

\section{Introduction}
Quantum networks could facilitate the development of quantum applications that are infeasible on classical networks (e.g., blind quantum computing (BQC) \cite{BroadbentEtAl2009}, distributed quantum computing (DQC) \cite{CaleffiEtAl2024}, or quantum key distribution (QKD) \cite{GisinEtAl2002, ScaraniEtAl2009}) and may ultimately enable a quantum internet \cite{Kimble2008, Wehner2018, KumarEtAl2025}.
The primary role of quantum networks is to distribute \ac{E2E} entanglement between distant nodes \cite{VanMeter2014}, thereby supporting the realization of quantum applications. In such a network, nodes can communicate by transmitting quantum bits (qubits) \cite{WeiEtAl2022}. These basic units of quantum information can be extended over long distances using a fundamental building block: quantum repeaters \cite{BriegelEtAl1998, AzumaEtAl2023}.
In this widely adopted approach, multiple hops are required to transmit quantum information between distant nodes. Entanglement is first created with elementary links and then extended through entanglement swapping to obtain an \ac{E2E} entangled pair \cite{ZukowskiEtAl1993, VanMeter2014}.
However, in practice, quantum networks operate with fragile particles that impose additional constraints (e.g., decoherence, fidelity, and the no-cloning theorem) \cite{WoottersEtAl1982, CacciapuotiEtAl2020}.
Specifically, entanglement generation and swapping operations are stochastic, with finite coherence times and limited quantum memory for entanglement storage/lifetime \cite{JulsgaardEtAl2004, LiuEtAl2021, SangouardEtAl2011}.
Consequently, scheduling and distribution of \ac{E2E} entangled pairs with multiple users in a quantum network requires a scheduler \cite{DahlbergEtAl2019, CicconettiEtAl2021} that answers these questions: when, for how long, and which resources should be involved to provide \ac{E2E} entangled pairs to serve the users' requests while accounting for the quantum constraints.

Recent progress has led to the development of an operating system for quantum network nodes (QNodesOS) \cite{DelleDonneEtAl2025} and an execution environment (Qoala) \cite{VanDerVechtEtAl2025}, which support multitasking and optimize the local use of entanglement pairs. Building on these abstractions, a recent \textit{generate-when-requested} architecture introduced the notion of \ac{PGA}, which is an allocation time along a selected path to attempt to produce a packet of entanglement, which succeeds with a target probability \(p_{packet}\) set by the network \cite{BeauchampEtAl2025}. This probability is a key architectural design parameter for network resource provisioning under a finite time budget. Indeed, due to the stochastic nature of quantum processes, no finite \ac{PGA} duration can provide a genuine guarantee of success; it would only be possible with an infinite allocation time \cite{BeauchampEtAl2025}. This implies an inherent trade-off between reliability and allocation budget of time/resources: targeting a higher \(p_{packet}\) requires a larger budget, whereas targeting a lower \(p_{packet}\) reduces the budget, but entails a higher probability of failure \(1-p_{packet}\). An objective of our work is to assess the performance sensitivity to this networking trade-off under contention.

In addition, the architecture presented must also determine how to schedule and distribute these \acp{PGA}. The authors proposed a central controller that periodically computes a static schedule of \acp{PGA} using the well-studied \ac{EDF} algorithm from real-time systems \cite{LiuEtAl1973}. From a systems perspective, this approach provides predictability that simplifies admission/coordination and operational stability, with fixed-interval computation/limited disruptions from nodes to the controller. However, at the cost of reduced responsiveness, notably to stochastic outcomes and asynchronous arrivals. Another objective is to evaluate, within this architecture, the performance impacts of runtime reactivity. Our main contributions are:
\begin{itemize}
    \item We propose a dynamic, online \ac{EDF}-like scheduler that reacts to arrivals and PGA outcomes, making runtime decisions (admit/schedule/defer/retry/drop) to resolve contention while prioritizing the earliest deadlines.
    \item We quantify the runtime reactivity benefits by (i) comparing a static baseline based on \ac{EDF} with the proposed dynamic scheduler and (ii) characterizing the dynamic scheduler's scalability as \(p_{packet}\) varies, reporting makespan, completion ratio, throughput, and link utilization/waiting time.
    \item We provide a model to map a target reliability \(p_{packet}\) to a PGA time budget under a common slot-based packet-generation model.
\end{itemize}

The rest of the paper is structured as follows.
In \Cref{sec:related}, we briefly review recent studies that are more relevant to our work.
In \Cref{sec:model}, we introduce the system model, while the main contribution is illustrated in \Cref{sec:eps}, which describes static vs.\ dynamic versions of a \ac{PGA} scheduling algorithm.
We evaluate the performance with simulations in \Cref{sec:eval}.
Finally, \Cref{sec:conclusion} summarizes our main findings.

\section{Related Work}\label{sec:related}
Quantum networks with multiple users have motivated studies on \ac{E2E} scheduling and distribution, in which a scheduler must account for stochastic operations (entanglement generation and swapping) and limited quantum memories.

Skrzypczyk and Wehner \cite{SkrzypczykWehner2021} introduced a centralized TDMA quantum network architecture that orchestrates entanglement delivery. Along similar lines, Beauchamp et al. \cite{BeauchampEtAl2025} presented a modular on-demand architecture with \acp{PGA}. This paper is the primary reference for our work. We develop two extensions of their architecture: (i) an online, event-driven scheduler that adapts \ac{PGA} decisions to request arrivals and completion outcomes, and (ii) a network-layer model that captures per-slot link-generation and probabilistic entanglement swapping.

Prior work \cite{CicconettiEtAl2021, HuangEtAl2025, NiEtAl2025} has proposed dynamic scheduling that assigns requests at the slot level. In contrast, our scheduler makes runtime decisions at the \ac{PGA} level by allocating a contiguous block of slots in response to each application's request. Gu et al. \cite{GuEtAl2023} have also developed a general optimization framework for entanglement scheduling/distribution (EDSI), in which each link continuously attempts to generate entangled pairs in a buffered network. We instead assume that entanglement generation is triggered only upon the arrival of a request, and that no pre-generated entangled pairs are available beforehand.

Finally, Lyapunov drift and Max-Weight queueing approaches \cite{VasantamEtAl2022, FittipaldiEtAl2022} provide analytical frameworks for contention and scheduling with respect to stability and throughput. These works propose mathematical frameworks under assumptions that are necessary to keep the models tractable. Our work takes a different approach: we present a practical setup for implementing an entanglement distribution network that could, in the future, benefit from integration with these theoretical optimization frameworks.

\section{System Model} \label{sec:model}

In this section, we present the network-layer system model used to schedule entanglement packets across a quantum network to meet deadlines. We provide the notations in \Cref{tab:notation}.

\begin{table}[!tbp]
    \caption{Notation}
    \label{tab:notation}
    \centering
    \scriptsize
    \begin{tabular}{c|c}
    \textbf{Symbol} & \textbf{Definition} \\
    \hline
        \(\mathcal{A}\)  & set of active applications \\
        \(a\)  & application index \\
        \(k\)  & \ac{PGA} index \\
        \(I_a\)  & number of entanglement packets over the session for application \(a\)  \\
        \(q_a\)  & number of \ac{E2E} entangled pairs for application \(a\) \\
        \(r_{a,k}\)  & release time of the \(k\)-th \ac{PGA} of application \(a\) \\
        \(D_{a,k}\)  & absolute deadline of the \(k\)-th \ac{PGA} of application \(a\) \\
        \(T_{a}\)  & period of application \(a\) \\
        \(H_i\)  & hyper-period \\
        \(\tau\) & slot duration \\
        \(Q_e\) & event queue\\
        \(Q_r\) & ready queue \\
        \(\pi_a\)  & path assigned to application \(a\) \\
        \(E(\pi_a)\)  & set of edges in the path of \(\pi_a\)\\
        \(L\)  & number of elementary links on \(\pi_a\) \\
        \(m\) & number of independent link-generation trials per slot \\
        \(p_{gen}\) & success probability of a single link-generation \\
        \(p_{link}\) & per-slot probability success elementary link \\
        \(p_{bsm}\) & \ac{BSM} success probability \\
        \(p_{e2e}\) & per-slot \ac{E2E} success probability (approximation)\\
        \(p_{packet}\)  & target probability that a \ac{PGA} generates at least \(q_a\) \ac{E2E} pairs \\
        \(B_{a}\) & allocated time budget of \ac{PGA} of application \(a\) \\
    \end{tabular}
\end{table}

\begin{figure}[!tbp]
    \centering
    \includegraphics[width=0.6\columnwidth]{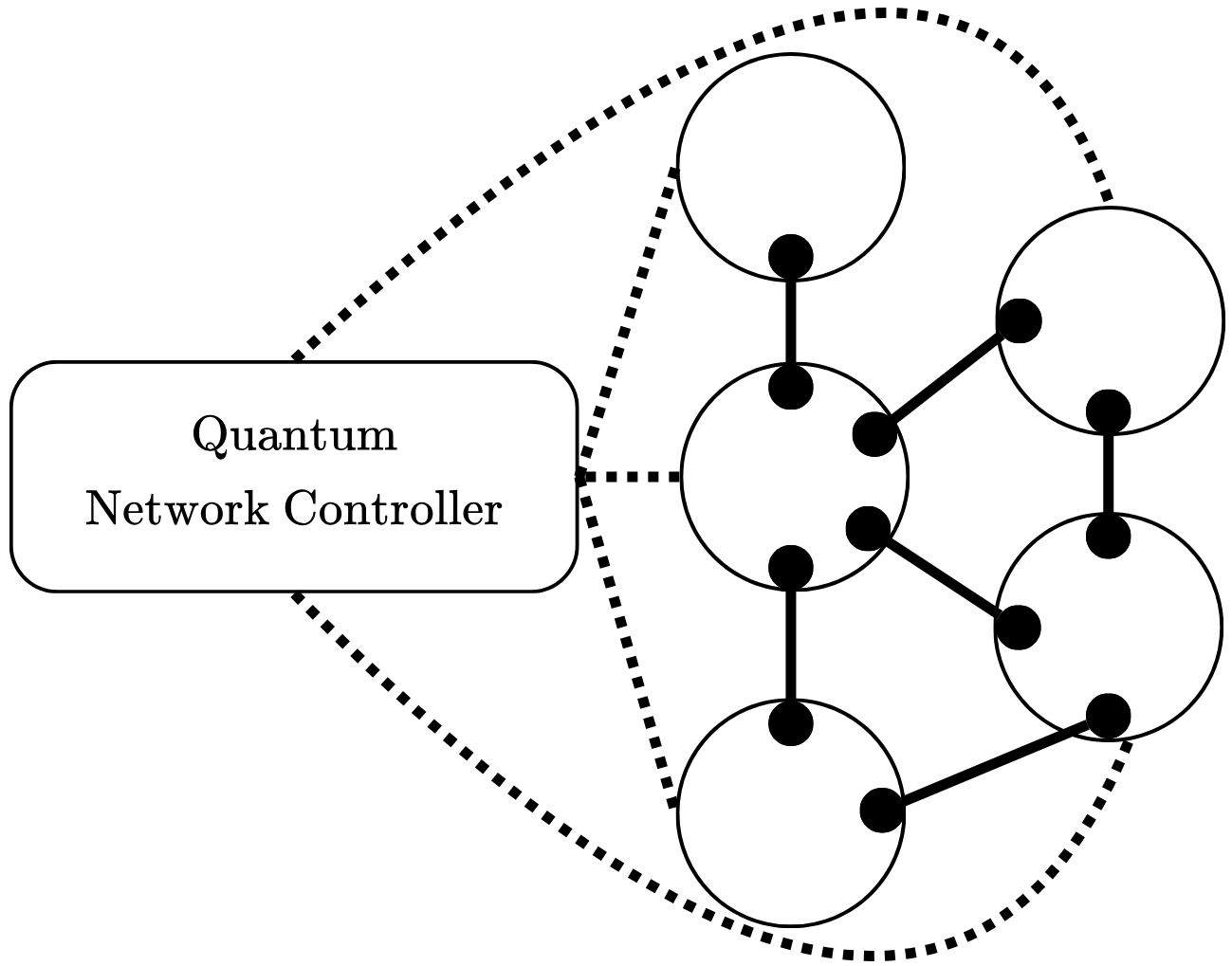}
    \caption{A quantum network with a central controller that coordinates 5 nodes. Each has a certain number of quantum links.}
    \label{fig:qn}
\end{figure}

\subsection{Quantum Network On-demand}\label{ssec:model:network}

First, we consider a quantum network, represented by an undirected graph, \(G = (V, E)\) where each node \(v \in V\) characterizes a quantum component that is capable of generating a bipartite entanglement with its neighbors. Each edge \(e \in E\), called a quantum link, indicates whether a direct entanglement generation is possible. Consequently, each node \(v \in V\) has at least one communication qubit per adjacent link, as illustrated in \Cref{fig:qn}. For simplicity, we do not consider multipartite entanglement \cite{NainEtAl2020}. In addition, we assume an on-demand entanglement distribution: the entanglement procedure begins only upon a user's request. We adopt the architecture described in \cite{BeauchampEtAl2025}, in which the central controller schedules entanglement packet requests across the network to maximize the number of requests fulfilled, using \acp{PGA} defined in \Cref{subsec:pga}.

An application \(a\) is characterized by a source-destination pair \((v^{\mathrm{src}}_{a}, v^{\mathrm{dst}}_{a})\), and requires the delivery of \(I_a\) packets of entanglement over a session. We also define a packet of entanglement consisting of \(q_a\) \ac{E2E} entangled pairs that are available simultaneously. The application has an initial arrival \(r_a^0\), and requests one packet every period \(T_a\). In this work, the central controller selects simply the shortest path \(\pi_a\) between \((v^{\mathrm{src}}_{a}, v^{\mathrm{dst}}_{a})\), using Dijkstra's algorithm. We have not addressed the quantum routing problem, which has been well studied in the literature \cite{Caleffi2017, Pant2019}. We leave for future work the investigation of a possible joint routing/scheduling with entanglement packet scheduling.

In addition, to handle contention between the applications on the link resources, we define a conflict graph \(G_c=(\mathcal{A, C})\), over active applications \(\mathcal{A}\), where \((a,b)\in \mathcal{C}\) if and only if the sets of links used by the applications overlap, i.e., \(E(\pi_a) \cap E(\pi_b) \neq \varnothing\). Thus, the scheduler may schedule any set of applications that forms an independent set in \(G_c\), i.e., allocates non-overlapping time windows to conflicting paths while allowing parallel execution for non-conflicting paths. 

\subsection{End-to-End Entanglement Distribution} \label{subsec:e2e}
We consider a time-slotted model in which the network controller operates in discrete slots of duration \(\tau\). Within each slot, each elementary quantum link can execute up to \(m\in\mathbb{N}\) independent entanglement-generation (e.g., enabled by temporal/frequency multiplexing in quantum memories \cite{CollinsEtAl2007, PuEtAl2017, YangEtAl2018}), each modeled as a Bernoulli trial with success probability \(p_{gen} \in (0,1)\). The success probability that an elementary link \(p_{link}\) has established at least one heralded entangled pair within a slot is
\begin{equation}
    p_{link} = 1 - (1-p_{gen})^{m}.
\end{equation}

Consider an application \(a\) assigned a path \(\pi_a\) comprising \(L\) elementary links and thus requiring \(L-1\) swaps. We assume that entanglement swapping is performed once all the elementary links are available, and each \ac{BSM} succeeds independently with probability \(p_{bsm} \in (0,1)\). 
The approximation success probability \(p_{e2e}\) of generating one \ac{E2E} entanglement pair is
\begin{equation}
        p_{e2e} \approx (p_{link})^{L} \cdot (p_{bsm})^{L-1}.
\end{equation}

This approximation abstracts away physical-layer correlations and detailed timing effects, yielding a tractable service model that schedulers can use to study contentions at the network layer. We discuss the sensitivity of our conclusions to these assumptions in \Cref{subsec:discussion}.

\subsection{Packet Generation Attempts (PGAs)} \label{subsec:pga}

A \ac{PGA} is a finite time budget allocated by the network to answer an application's request \(a\) by attempting to produce one entanglement packet (\(q_a\) entangled pairs in the allocated budget); the application requires \(I_a\) packets over the session. An application \(a\) is thus defined by a series of \acp{PGA} \(k \in \mathbb{N}\). The \(k\)-th PGA of application \(a\) is released at time \(r_{a,k} = r_a^0 + kT_a\), and has an absolute deadline \(D_{a,k} = r_{a,k} + T_a\). \(\text{PGA}_{a,k}\) is released periodically every \(T_a\), and each must be completed before \(D_{a,k}\). A \ac{PGA} reserves the required path resources and is non-preemptive, in the sense that it cannot be suspended and resumed, but supports early completion. If a \ac{PGA} accumulates \(q_a\) \ac{E2E} entangled pairs before its allocated duration expires then:
\begin{itemize}
    \item the \(\text{PGA}_{a,k}\) is considered successful (or completed), i.e., one packet of entanglement has been delivered. If \(I_a\) is reached, the application \(a\) is considered fully served.
    \item the reserved resources are freed immediately.
\end{itemize}

We now define the duration of a \ac{PGA} \(B_a\) as the time budget to generate \(q_a\) \ac{E2E} entangled pairs with at least probability \(p_{packet}\). Let $n$ be the number of slots allocated to the PGA. We treat each slot as one independent \ac{E2E} generation trial with probability of success \(p_{e2e}\). The number of $q_a$ \ac{E2E} pairs generated during the PGA satisfies \( X \sim \text{Binomial}(n, p_{e2e})\), and we require 
\begin{equation}
    \Pr[X\ge q_a] = \sum_{x=q_a}^{n} \binom{n}{x} p_{e2e}^{x}(1-p_{e2e})^{n-x} \ge p_{packet}.
\end{equation} We choose the minimum number of slots
\begin{equation}
    n'=\min \left\{ n \in \mathbb{N} : \Pr[X \ge q_a] \ge p_{\text{packet}}\right\}.
\end{equation}
We then set the PGA duration $B_a$ as:
\begin{equation}\label{eq:duration}
    B_a = n'\tau.
\end{equation}We treat deadline feasibility as a hard constraint: $B_a \leq T_a$, i.e., if $B_a$ exceeds the period \(T_a\), the application's request is infeasible. We assume that once an \ac{E2E} pair is generated during a \ac{PGA}, it can be stored without decoherence until the end of $B_a$. However, unused pairs are discarded at the end of a \ac{PGA}; therefore, \ac{E2E} entangled pairs are not carried over across \acp{PGA}. This buffering assumption abstracts away memory imperfections and isolates the impact of network-layer runtime scheduling decisions. We note that the assumptions in this section differ from those in \cite{BeauchampEtAl2025}, which evaluates the architecture in the absence of quantum memories and without exploiting the early termination of \acp{PGA}.

\begin{figure*}[!t]
    \centering
    \includegraphics[width=\textwidth]{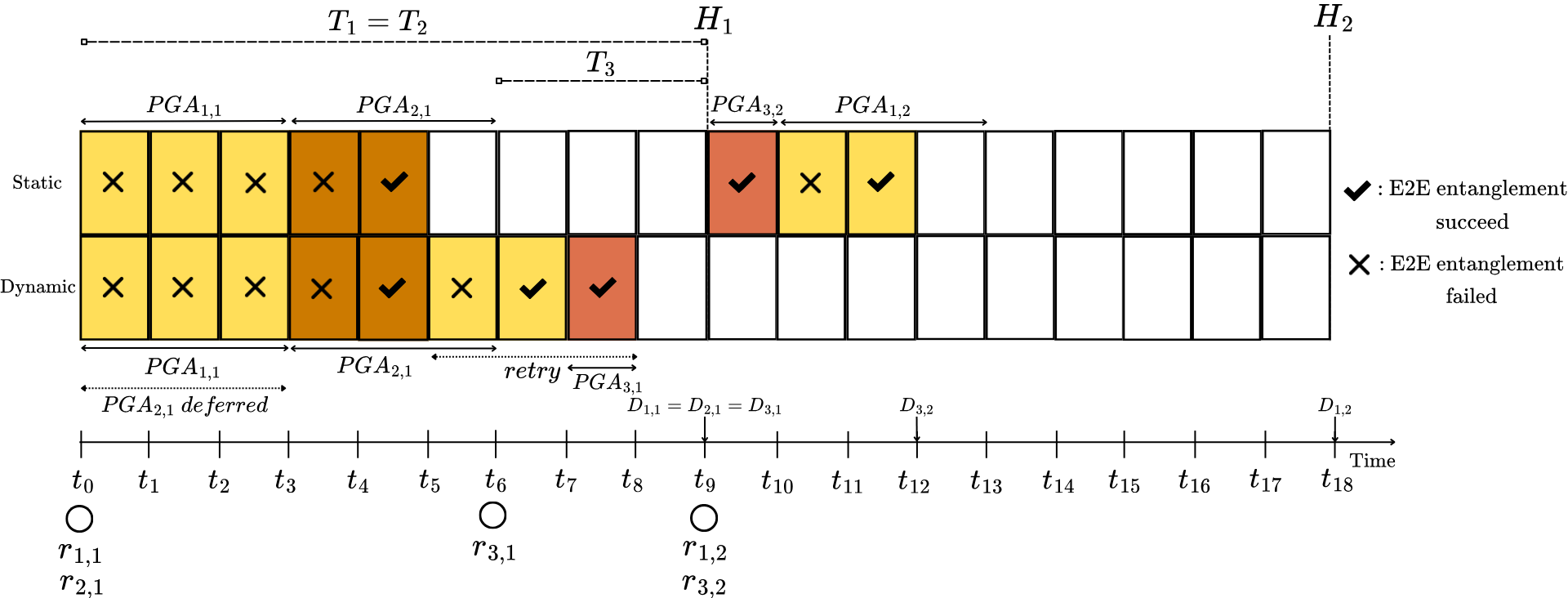}
    \caption{A toy example comparing the static (top) and dynamic (bottom) schedulers, where 3 applications \(a\) request one entanglement packet (\(\forall a\in \mathcal{A},\; I_a = 1\)) with periods \(T_a\). 2 applications arrive at \(t_0\), while the last arrives at \(t_6\). Each scheduler schedules \(\text{PGA}_{a,k}\) (time budgets), represented by solid double arrows. In each slot, an attempt of \(E2E\) entanglement is performed (entanglement generation and swapping with all the links of the selected path \(\pi_a\)). Each color is associated with an application and an outcome represented by a symbol. The hyper-periods \(H_i\) are used only by the static scheduler. The dashed double arrows indicate specific runtime mechanisms of the dynamic scheduler (defer and retry). We observe that, in the static case, the last application, \(a=3\) (in red), misses its deadline at \(t_9\). For the dynamic case, it is scheduled using the early-completion feedback at \(t_7\), when all links of \(\pi_a\) are available.}
    \label{fig:pga}
\end{figure*}

\section{Entanglement Packet Scheduling}\label{sec:eps}
In this section, we first formulate the entanglement packet scheduling in \Cref{subsec:pb}. We then present our baseline, a static schedule computed using an \ac{EDF}-based approach to generate periodic timetables for \acp{PGA}, described in \Cref{subsec:static}.
Then, in \Cref{subsec:dynamic}, we propose an online scheduler that makes decisions based on the current conditions of the assignment of \ac{PGA} to the nodes.

\subsection{Entanglement Packet Scheduling} \label{subsec:pb}

For each slot \(t\), the scheduler selects a set of \acp{PGA} to start, 
\begin{equation}
    S(t) \subseteq \{(a,k): r_{a,k} \leq t\},
\end{equation}
where starting \(\text{PGA}_{a,k}\) at slot \(t\) reserves all links on its assigned path \(\pi_a\), i.e., the edge set \(E(\pi_a) \subseteq E\). Let \(R(t)\subseteq E\) be the set of links already reserved at the beginning of slot \(t\) by previous \acp{PGA} that are not completed. The decision must satisfy
\begin{align}
E(\pi_a) \cap R(t) &= \varnothing, && \forall (a,k) \in S(t) \\
E(\pi_a) \cap E(\pi_{a'}) &= \varnothing, && \forall (a,k) \neq (a',k') \in S(t).
\end{align}
Let \(P_a(t)\) be the number successfully completed \acp{PGA} (or packets of entanglement) of application \(a\) up to slot $t$, and \(I_a\) the required number. We define
\begin{equation}
    \mathcal{A} (t)= \{a:P_a(t) < I_a\}.
\end{equation}
The end of the session is the slot \(t_{end}\) where all applications have been served, i.e., \(\forall_a, P_a(t_{end})=I_a\). Thus, the scheduler's objective is to continue scheduling until \(\mathcal{A}(t) = \varnothing\).

\subsection{Baseline: Static EDF-based Scheduler of PGAs} \label{subsec:static}
We consider a central controller that periodically computes a static schedule (offline) for non-preemptive \acp{PGA} over a fixed planning horizon. The scheduler calculates the hyper-period \(H_i\), the \ac{LCM} of the periods of the applications, and the \ac{PGA} duration for each application using \Cref{eq:duration}. Within \(H_i\), the controller schedules the \acp{PGA} using an \ac{EDF} policy. The scheduler maintains a conflict graph, as defined in \Cref{sec:model}, which enables updating the set of parallel \acp{PGA} at each \(H_i\). Specifically, multiple \acp{PGA} can be executed concurrently, depending on link/resource availability (e.g., two \acp{PGA} that require the same quantum link cannot begin entanglement processes simultaneously). Admission control is necessary before a schedule is distributed to the nodes. A schedule is considered feasible if all \acp{PGA} can be executed before their deadlines, as determined by the conflict graph. Otherwise, a schedule is considered infeasible.

A toy example is shown in \Cref{fig:pga} and illustrates how the static scheduler operates. First, it computes a hyper-period \(H_1\) with the set of active applications \(\mathcal{A}(t=0) =\{1, 2\}\), which is \(H_1 = \text{LCM}(T_1,T_2) = 9\). For simplicity, in this example, concurrency is not possible, so the schedulers can schedule at most one \ac{PGA} at a time. The schedule at $H_1$ is feasible; thus, the controller schedule according to an \ac{EDF} policy. In this case, both \acp{PGA} have the same deadline (the choice is then arbitrary, since they have the same arrival time). The scheduler then recomputes the schedule for the next hyper-period, \(H_2 = \text{LCM}(T_3, T_1) = 9\), and repeats the process for each hyper-period until all applications have received their specified number of entanglement packets \(I\). Consequently, the static scheduler has not scheduled a \ac{PGA} for the application \(3\) that arrives at \(t_{6}\), where the deadline has been missed at $t_9$. Another consequence is that the use of early success is not possible, notably at \(t_5\) and \(t_{12}\). All applications have been served at \(t_{13}\).

\subsection{Dynamic Scheduler of PGAs}\label{subsec:dynamic}
The static scheduler above produced a timetable of \acp{PGA} for a set of applications.  
Instead, the dynamic scheduler reacts to the status of new and current applications (called events) to assign \acp{PGA} to nodes. We consider the following types of events: (i) the arrival of a request (a new request or the release of a new PGA for the same request) with possible deferrals and (ii) the PGA completion outcomes (success or failure) with possible retries. To manage these events, the dynamic scheduler maintains two queues: requests wait in the event queue $Q_e$ until the ready queue $Q_r$ processes them.

At each scheduling decision point, the scheduler may admit application's request and schedule multiple \acp{PGA} (at most one per application request $a$) based on link availability and path concurrency, i.e., two requests can be scheduled simultaneously if and only if the paths selected by the controller do not share a link, see \Cref{ssec:model:network}. 
As in the static version, the scheduler computes the budget time $B_a$ from \cref{eq:duration}, schedules \acp{PGA} under the \ac{EDF} policy, and breaks ties by the earliest release time. 
Based on the current state of the network, the scheduler makes different decisions.
If the link resources are available and the PGA meets its deadline, the PGA is scheduled immediately. If the PGA fails but there is still time before the deadline and resources are available, it is rescheduled with the same deadline/duration (i.e., retried); otherwise, it is dropped. If the requested resources are currently unavailable but the \ac{PGA} may still meet its deadline, then the \ac{PGA} is deferred to a later start time once the resources become available. The pseudo-code of the dynamic scheduler is reported in \Cref{alg:dynamic}. 
Regarding the overhead, the dynamic scheduler updates the priority queues $Q_e$ and $Q_r$ and checks the resource availability of the selected path $\pi_a$. The queue updates cost, respectively, $O(\log n_e)$, where $n_e$ is the number of elements in $Q_e$, and $O(\log n_r)$, where $n_r$ is the number of elements in $Q_r$. Checking the path cost takes $O(|\pi_a|)$, where $|\pi_a|$ is the number of links in the selected path. Thus, the total cost per-event is $O(\log n_e + \log n_r + |\pi_a|)$. In the worst case under saturation, the event rates (deferrals and retries) are bounded because once a \ac{PGA} can no longer meet its deadline, the scheduler drops it, preventing starvation.

\begin{algorithm}[!t]
\caption{Dynamic EDF-like Scheduling of \acp{PGA}}
\label{alg:dynamic}
\scriptsize
\textbf{Input:} \textit{EventQueue} (ordered by event times), \textit{ReadyQueue} (ordered by earliest deadline)
\begin{algorithmic}[1]
\While{there exists a PGA either scheduled as an event or in \textit{ReadyQueue}}

    \If{\textit{ReadyQueue} has no PGA}
        \State wait until the next event time in \textit{EventQueue}
    \EndIf
    
    \State move all due events from \textit{EventQueue} into \textit{ReadyQueue}
    \While{\textit{ReadyQueue} has a PGA}
        \State select the PGA with the earliest deadline in \textit{ReadyQueue}
        \If{PGA cannot meet its deadline}
            \State drop PGA
        \ElsIf{PGA cannot start because required links are busy}
            \State defer PGA into \textit{EventQueue} at the earliest time its links are free
        \Else
            \State attempt PGA
            \If{PGA succeeds}
                \State update the remaining number of required PGAs for the application
            \ElsIf{PGA fails and there is enough time before its deadline}
                \State retry PGA into \textit{EventQueue} at attempt end (same duration/deadline)
            \Else
                \State drop PGA
            \EndIf
        \EndIf
    \EndWhile
\EndWhile

\end{algorithmic}
\end{algorithm}

In the toy example shown in \Cref{fig:pga}, we illustrate several mechanisms by which the dynamic scheduler operates. First, the scheduler responds at \(t_0\) due to the arrival of the two applications, with the set of active applications \(\mathcal{A}(t=0) =\{1, 2\}\), enqueued in $Q_e$. Since resources are available, \ac{PGA} with the earliest deadline $PGA_{1,1}$ (arbitrary choice in this case) is processed into $Q_r$ and scheduled at $t_0$. $PGA_{2,1}$ is deferred to the earliest time the links are available ($t_3$), and waits in $Q_e$ to be processed again. The $PGA_{1,1}$ failed; however, there is enough time ($B_1 \leq D_{1,1}$) before its deadline. Consequently, a retry is rescheduled for $PGA_{1,1}$ (placed in $Q_e$) at the end time of the attempted \ac{PGA} ($t_3$), with the same initial budget time \(B_1 = 3\) and the same deadline $D_{1,1}$. The $PGA_{2,1}$ is then processed and succeeds earlier, the retry of $\text{PGA}_{1,1}$ is scheduled directly at \(t_5\). Similarly, the early-completion feedback is exploited at \(t_7\) by the dynamic scheduler, which schedules the last-arrival application with \(PGA_{3,1}\), thus meeting the deadline \(D_{3,1}\). All applications have been served at \(t_{8}\).

\section{Performance Evaluation}\label{sec:eval}
In this section, we present the potential benefits of a dynamic entanglement packet scheduler for the on-demand architecture. To do so, we compare the static vs.\ dynamic schedulers evaluated on the backbone of the Italian research and education network, which is operated by the GARR consortium. We have chosen this topology because it comprises a small set of high-degree nodes (e.g., up to 10 neighboring nodes), thereby creating bottlenecks and contention, and because it has also been used in other work on admission control for quantum networks \cite{CicconettiEtAl2022}. We have then assessed the dynamic approach under increasing application load and dynamic arrivals.

\subsection{Experimental Setup} \label{subsec:exp}
To obtain the results presented in this section, we implemented an open-source Python simulator \cite{Tran2026}. We have evaluated our experiments on the GARR topology that comprises 48 nodes and 62 links. Additionally, based on prior work \cite{DahlbergEtAl2019, SkrzypczykWehner2021, BeauchampEtAl2025}, we use the parameters provided in \Cref{tab:sim}. Each application requests $I_a=100$ entanglement packets per session; each packet contains $q_a= 2$ \ac{E2E} entangled pairs, with period \(T_a=1s\). Each application request is associated with randomly selected source and destination nodes. In \Cref{tab:garr}, the probability of observing a certain number of hops for the GARR network is shown. We select paths between nodes using Dijkstra's shortest path algorithm. All physical-layer and application parameters are fixed to the values in \Cref{tab:sim}. The evaluation varies only in the number of concurrently active applications and the target reliability $p_{packet}$.

\begin{table}[!t]
\caption{Garr network: probability distribution of observing a path length (in \# hops)}
\label{tab:garr}
\centering
\setlength{\tabcolsep}{3.5pt}
\begin{tabular}{|c|c|c|c|c|c|c|c|c|}
\hline
\textbf{\# Hops} & 1 & 2 & 3 & 4 & 5 & 6 & 7 & 8 \\
\hline
\textbf{Probability} & 0.055 & 0.157 & 0.285 & 0.282 & 0.155 & 0.051 & 0.013 & 0.002 \\
\hline
\end{tabular}
\end{table} 

\begin{table}[!t]
\caption{Simulation parameters}
\label{tab:sim}
\centering
\setlength{\tabcolsep}{3.5pt}
\begin{tabular}{|c|c|c|c|c|c|c|c|c|}
\hline
\textbf{Parameter}  &$\tau$ & $m$ & $T_a$ & $p_{gen}$ & $p_{bsm}$ & $q_a$ & $I_a$ & $\lambda$\\
\hline
\textbf{Value}  &$0.0001 ~\text{s}$ & $1000$ & $1~\text{s}$ & $0.001$ & $0.6$ & $2$ & $100$ & $1~\text{s}^{-1}$\\
\hline
\end{tabular}
\end{table}

To ensure comparable scheduling performance, we set the load to 50 applications; at this point, the static scheduler may reject some instances during admission. Consequently, for paired comparison, we report results only on instances admitted by the static scheduler (so both schedulers are evaluated on the same feasible set). Additionally, for the comparison, to isolate the scheduling performance under a fixed load from arrival-time variability, we consider that all application sessions are instantiated at \(t_0\) and remain active for the duration of the simulation horizon \([t_0,t_{end}]\).

In a second experiment, to study the scalability of the dynamic scheduler (only) under increasing loads and dynamic traffic, we generate sessions according to an exponential inter-arrival model with mean \(1/\lambda\) (i.e., a Poisson arrival process of rate \(\lambda\)), and a number of applications increasing from 50 to 300. For both evaluations, we have swept the values of \(p_{packet} = \{0.1, ..., 0.9\}\). 

The performance metrics we have used are: 
\begin{itemize}
    \item Admission rate is the number of feasible schedules divided by the total number of schedules.
    \item Completion ratio is the number of unique completed \acp{PGA} \(N_\text{completed}\) divided by the total number unique of \acp{PGA}, \(N_\text{total}\). Each successful \ac{PGA} corresponds to one delivered packet. Retries are internal rescheduling of the same \ac{PGA} instance and are not counted as additional \acp{PGA} in \(N_\text{total}\); a \ac{PGA} is counted as completed if any of its retries succeed.
    \item Makespan is the elapsed time from the first PGA arrival to the last PGA completion (when all applications have been served completely).
    \item \(\text{Throughput} = \frac{N_\text{completed}}{\text{makespan}}\). A retry is not inherently counted toward throughput, but it may cause an initial PGA to be counted as completed if one of its retries succeeds.
    \item The per-link utilization is the total time link \(i\) is busy divided by the makespan.
    \item The per-link waiting time is the total waiting time on link \(i\) divided by the number of PGAs that have used this link.
\end{itemize}

The results for each point were obtained by running 200 simulations with different random seeds, and the mean of each metric across runs was computed with the \(95 \%\) confidence interval. Specifically, for each seed, we evaluated both static and dynamic schedulers using the same random seed, following a ``paired run'' comparison methodology.

\subsection{Results} \label{subsec:results}

\begin{figure}[t]
    \centering
    \includegraphics[width=\columnwidth]{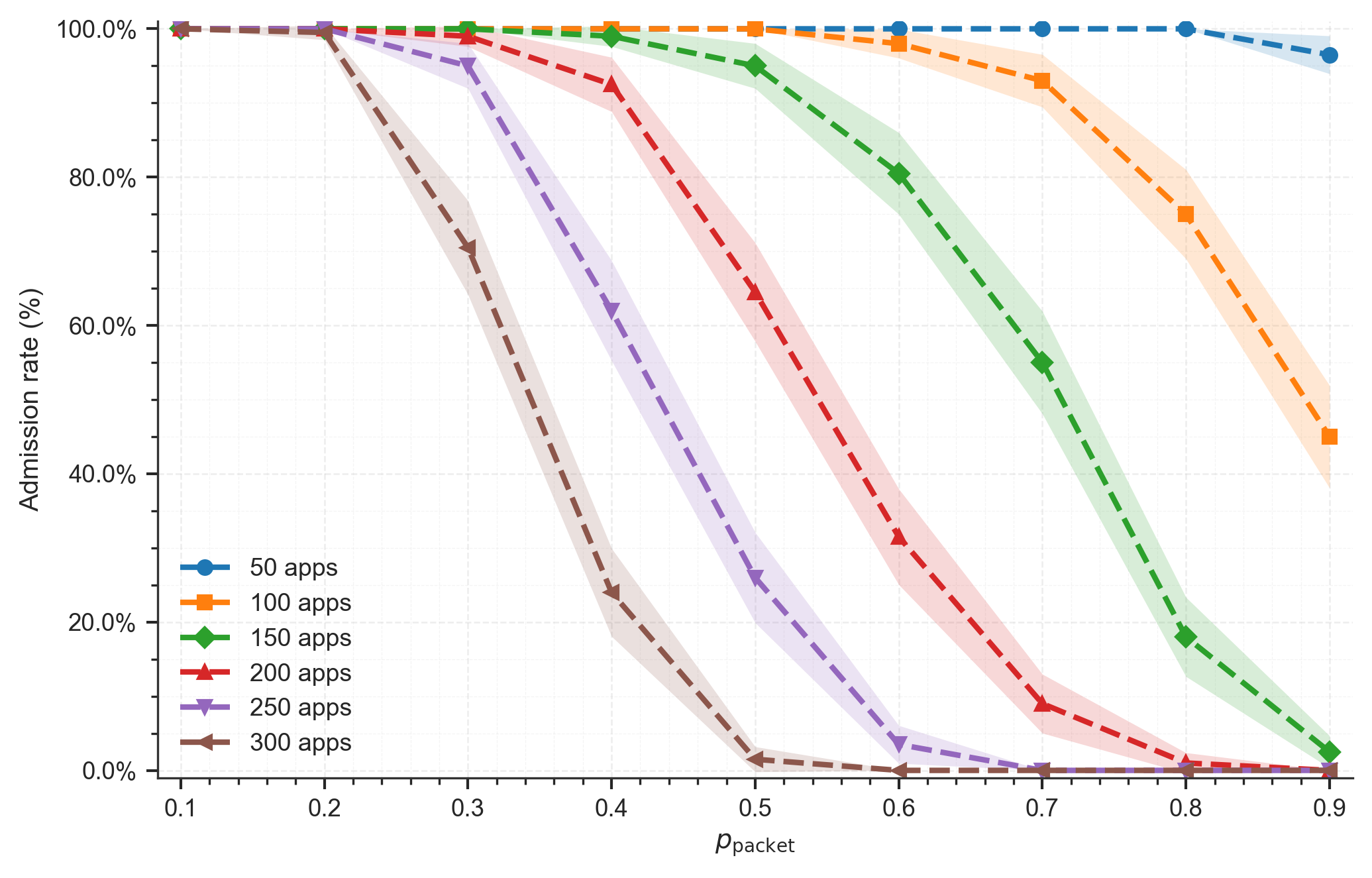}
    \caption{The static scheduler admission rate of feasible schedules vs. \(p_{packet}\).}
    \label{fig:admission}
\end{figure}

Using the same setup \Cref{subsec:exp}, we first evaluated the admission rate of the static schedule under the admission control in \Cref{subsec:static}. To do so, we have increased the number of applications based on different success probabilities set by the network \(p_{packet}\). The results, presented in \Cref{fig:admission}, show a correlation between the admission rate vs. \(p_{packet}\) and the number of applications. For small \(p_{packet} \leq 0.2\), the admission rate is near 100\% for all the loads. When \(p_{packet}\) increases along with an increasing load, we start to see a decrease in the admission rate. To a certain point, for \(p_{packet} \geq 0.6\) and applications \(\geq 250\), almost no schedules are admitted. This is due to high resource contention: complex schedules are rejected first. Our static scheduler computes a schedule in advance that produces a contention-free, feasible-deadline schedule using hyper-periods. As \(p_{packet}\) increases, \(B_a\) increases for each \ac{PGA}, and thus pushes the bottleneck links to their maximum capacity (and thus contention), which results in an abrupt drop in admission rate. This motivates an online scheduling approach that can defer and drop at runtime, rather than failing at admission.

\begin{figure*}[t]
    \centering
    \includegraphics[width=.42\textwidth]{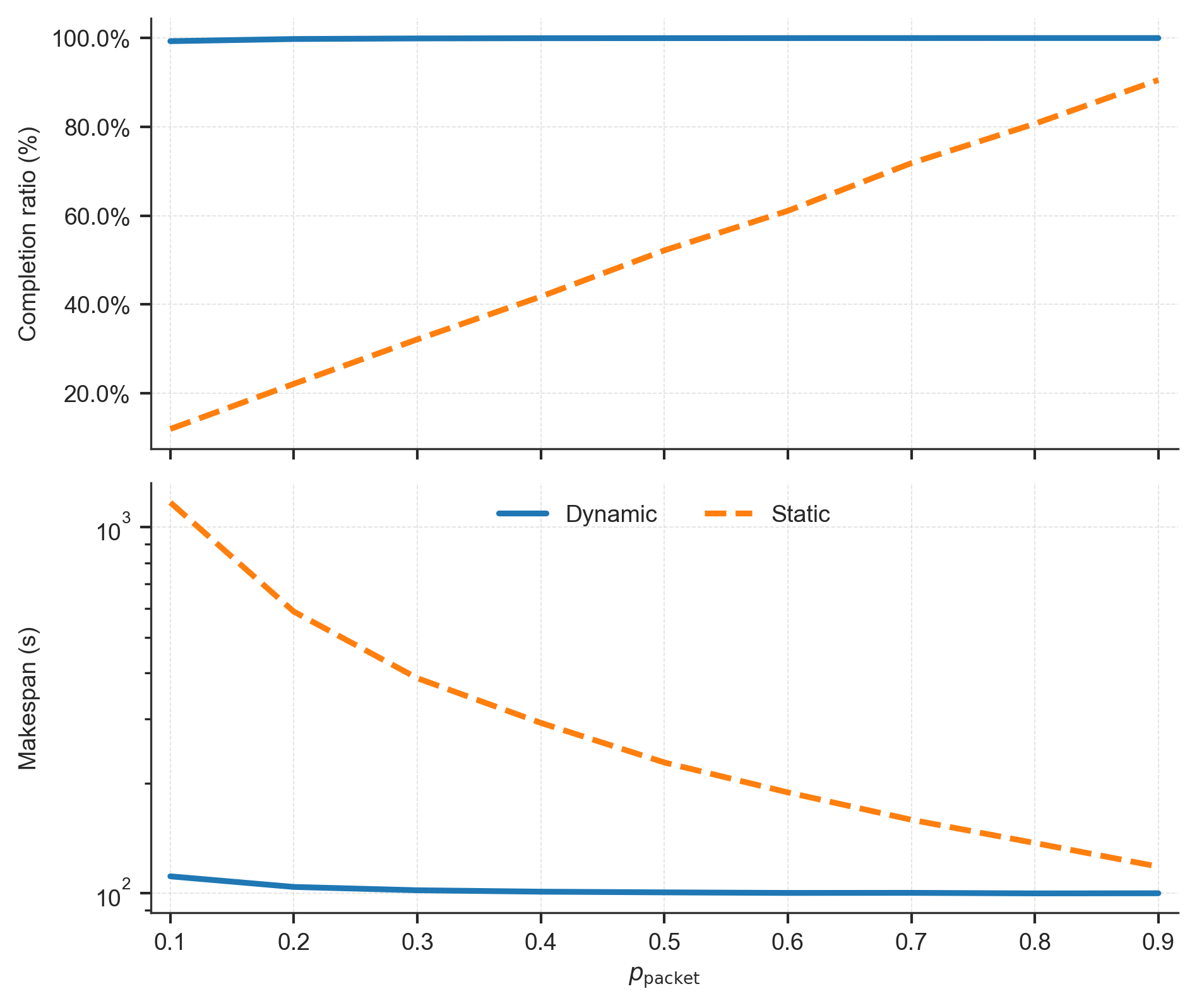}
    \includegraphics[width=.45\textwidth]{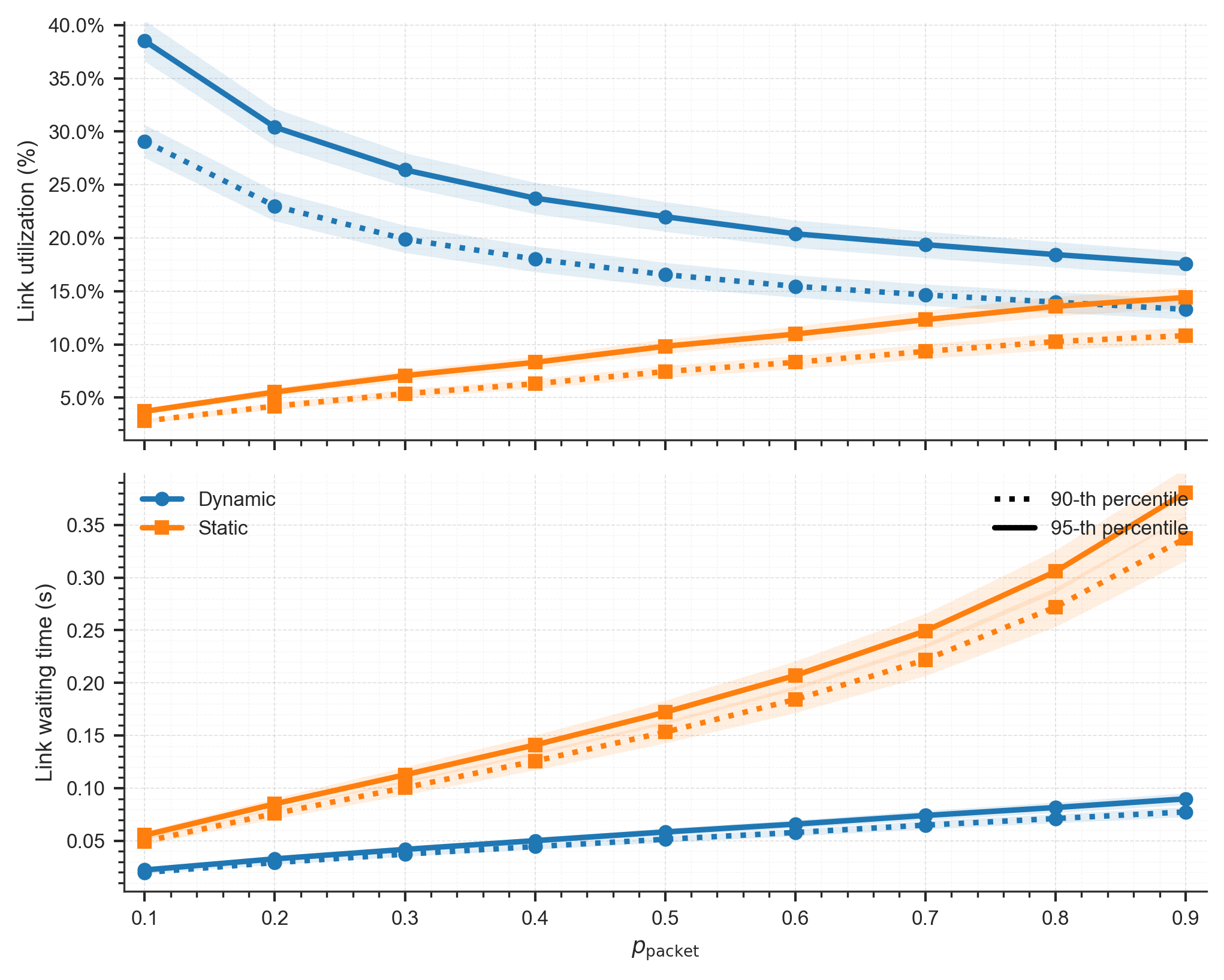}
    \caption{Comparison between the static and dynamic schedulers: the left figure shows the completion ratio (top) and makespan (bottom) in logarithmic scale vs. \(p_{packet}\); the right figure shows the 90th/95th percentile per-link utilization (top) and per-link waiting time (bottom)  vs. \(p_{packet}\).}
    \label{fig:50}
\end{figure*}

Secondly, we compared the static and dynamic schedulers with the setting described in \Cref{subsec:exp}. The results from this comparison are shown in \Cref{fig:50}. For all values of \(p_{packet}\), the dynamic scheduler shows the best performance. Because it can respond to stochastic outcomes: failures become retried when deadlines are feasible, \acp{PGA} are deferred until conflicting links are free, and early completions are used immediately.
Consequently, these mechanisms allow a completion ratio near \(100\%\), while keeping a makespan constant across \(p_{packet}\). For the static scheduler, at lower \(p_{packet}\), each \ac{PGA} is shorter and more likely to fail; it does not react to these failures and therefore misses many opportunities.
At higher \(p_{packet}\), the completion ratio, makespan, and link utilization results are almost the same. This is due to a changing regime in which the budget $B_a$ is expanded, successes take precedence over failures, and, consequently, the benefits of runtime are less pronounced. 
However, the link waiting time reveals how each scheduler achieves these results. With these longer budgets, each successful \ac{PGA} reserved resources for a longer time than low \(p_{packet}\).
This has a consequence, especially for bottleneck links: any other PGA that requires the same links must wait longer. Since the static has no runtime mechanism, increasing \(B_a\) mechanically increases the waiting times. In contrast, the dynamic scheduler defers \acp{PGA} that need a busy link and schedules other non-conflicting \acp{PGA}. In addition, when links are available (including early completion), the dynamic scheduler immediately assigns to the ready \acp{PGA} rather than paying the full price of the estimated allocated budget $B_a$.

\begin{figure*}[t]
    \centering
    \includegraphics[width=.52\textwidth]{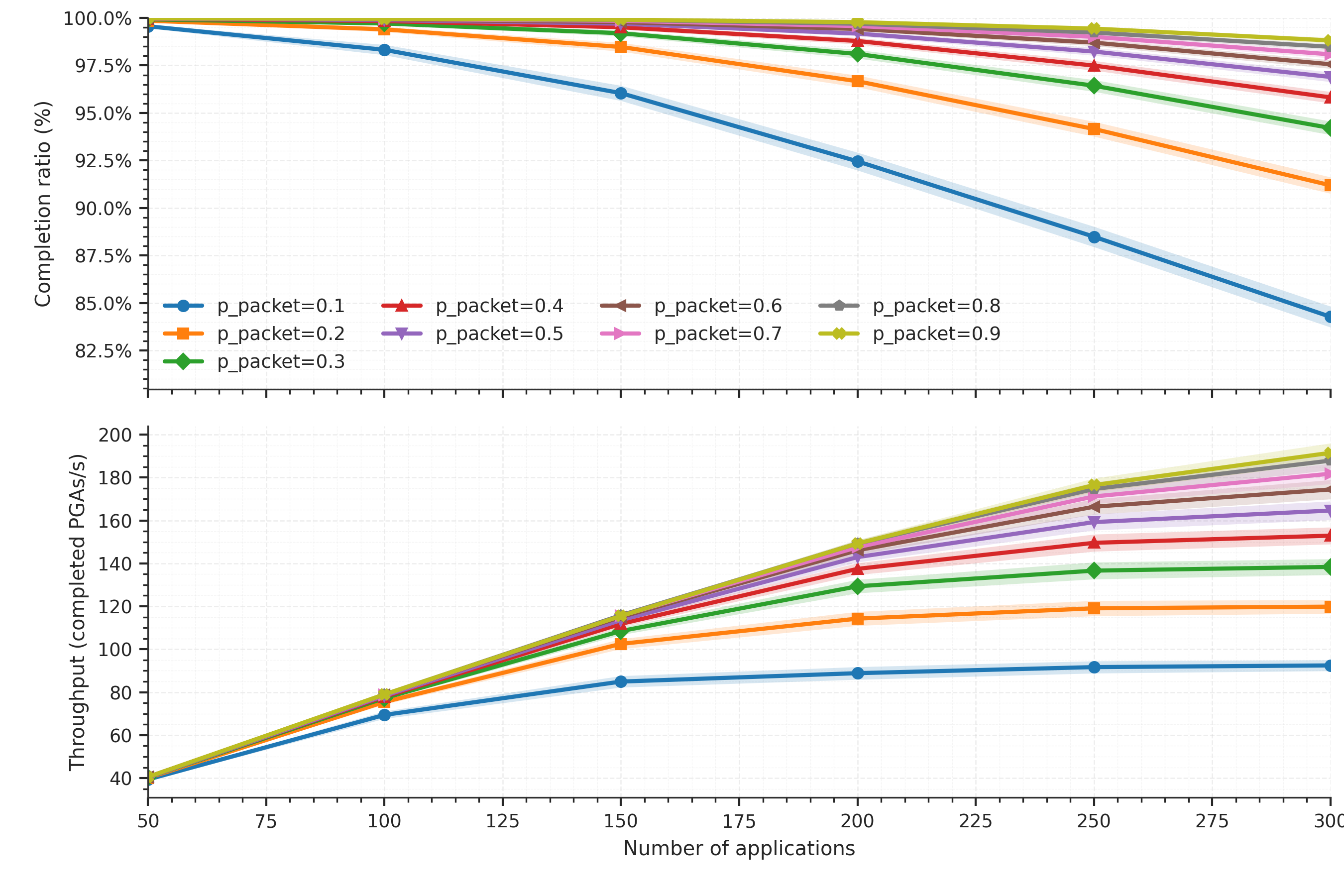}
    \includegraphics[width=.45\textwidth]{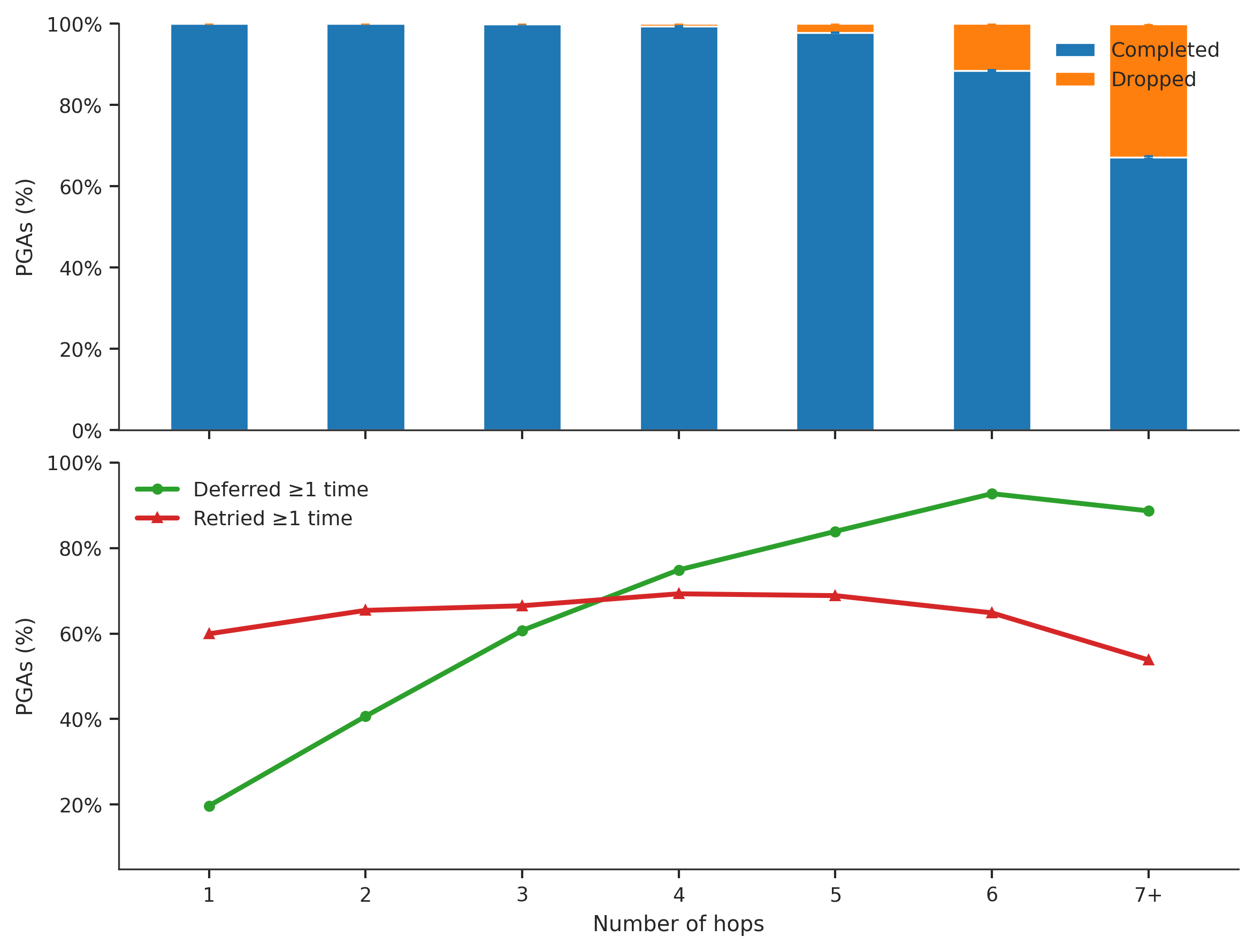}
    \caption{Performance on the load and the path length of the dynamic scheduler: the left figure shows the completion ratio and throughput according to an increasing load for different values of \(p_{packet}\); the right figure shows the proportion of \acp{PGA} completed/dropped and fractions of deferred/retried based on the number of hops (for \(n_{apps}=200,\  p_{packet} = 0.3\)). Note that a \ac{PGA} can be deferred and retried as long as it can meet its deadline.}
    \label{fig:dynamic}
\end{figure*}

Finally, we present the results of the dynamic scheduler, in \Cref{fig:dynamic}, under increasing load/dynamic traffic and its behavior at a specific medium/high contention point (200 applications and \(p_{packet}=0.3\), where we start to see a plateau for the throughput). When the load increases, the completion ratio stays high but starts to decrease (at worst \(84 \%\)), while throughput increases and is consistently higher with higher \(p_{packet}\). 
We observe at some point a plateau for throughput (which appears sooner for low \(p_{packet}\)), which underlines the saturation regime. The plateau occurs when a small set of high-degree nodes becomes saturated, so any additional load cannot increase the number of concurrent non-conflicting \acp{PGA} (and thus the throughput).
In the figure on the right, we see that the completion ratio stays at \(100 \%\) until 5 hops, then starts to deteriorate for longer paths (\(\approx 65\%\) for 7+ hops) with a proportional increase in drop. 
The percentage of deferred \acp{PGA} at least once increases with the number of hops (from \(\approx 20\%\) to \(\approx 90\%\)), while the retries remain almost constant but decrease at 7+ hops. Longer paths require more links simultaneously and, therefore, have a higher probability of encountering at least one busy link, which explains the monotonic increase of deferred \acp{PGA}. In addition, under saturation, the longest \acp{PGA} are more likely to miss their deadlines and are consequently dropped rather than deferred or retried, which penalizes longer-path requests; mitigating this bias via explicit fairness mechanisms is left for future work.

\subsection{Discussion and limitations} \label{subsec:discussion}
When the network starts to increase the reliability target \(p_{packet}\), the system allocates more resources to reduce stochastic failures, but at the cost of increasing the likelihood of contention.
Consequently, at higher \(p_{packet}\), bottleneck links saturate, and the static scheduler's admission rate drops sharply. In contrast, the dynamic scheduler makes runtime decisions: defer when blocked, retry when the deadline is feasible, drop when infeasible, and admit/schedule quickly using early-release feedback.
This suggests that an online scheduler with runtime decision-making capabilities could make more effective use of available resources. Moreover, the link utilization and plateaus indicate that under high contention, a small set of high-degree nodes saturates, confirming that scheduling/routing are tightly interconnected. A joint routing–scheduling could thus be an interesting direction (e.g., routing that is aware of scheduling/link congestion).
Additionally, the number of hops in this system appears to be a key performance parameter (i.e., longer paths are more likely to be dropped/deferred, require more resources, and have a lower completion ratio than shorter paths), motivating potential path-length-aware policies for admission control and/or service prioritization.

However, our system model has several simplifying assumptions. First, we have not accounted for the minimum fidelity required per application or for classical signal latencies, and we have abstracted elements in $p_{e2e}$ (instantaneous entanglement swapping and physical-layer details). Then, we have simplified the assumption about the quantum memory by assuming that an E2E entangled pair generated remains valid until the end of a PGA. Hence, our results may change with more realistic physical constraints. Nevertheless, in an ideal setup, our analysis shows the benefits of a scheduler that makes runtime decisions, which could serve as a benchmark for other studies. To further improve realism, we will extend the performance evaluation to account for heterogeneous application requests and link characteristics.

\section{Conclusion}\label{sec:conclusion}
In this paper, we presented a dynamic, online EDF-like scheduler that makes runtime decisions for entanglement packets and showed that this reactivity improves performance relative to the static baseline. These early results suggest that an on-demand entanglement packet architecture could benefit from an adaptive control layer that balances predictability and reactivity to scale under contention. They further suggest that co-design of scheduling and routing could be a promising direction for future work, along with improvements to the control plane.

\bibliographystyle{IEEEtran}
\bibliography{bib}

\end{document}